\documentclass{article}

\usepackage{microtype}
\usepackage{graphicx}
\usepackage{subfigure}
\usepackage{booktabs} 

\usepackage{hyperref}

\usepackage[accepted]{icml2023}

\usepackage{amsmath}
\usepackage{amssymb}
\usepackage{mathtools}
\usepackage{amsthm}

\usepackage[capitalize,noabbrev]{cleveref}

\theoremstyle{plain}
\newtheorem{theorem}{Theorem}[section]
\newtheorem{proposition}[theorem]{Proposition}
\newtheorem{lemma}[theorem]{Lemma}

\theoremstyle{definition}
\newtheorem{definition}[theorem]{Definition}
\newtheorem{assumption}[theorem]{Assumption}
\theoremstyle{remark}

\usepackage[textsize=tiny]{todonotes}

\begin{document}

\twocolumn[
\icmltitle{Representation Learning in Low-rank Slate-based Recommender Systems}

\begin{icmlauthorlist}
\icmlauthor{Yijia Dai}{yyy}
\icmlauthor{Wen Sun}{yyy}
\end{icmlauthorlist}

\icmlaffiliation{yyy}{Department of Computer Science, Cornell University, Ithaca, USA}
\icmlcorrespondingauthor{Yijia Dai}{yd73@cornell.edu}
\icmlcorrespondingauthor{Wen Sun}{ws455@cornell.edu}

\icmlkeywords{Machine Learning, ICML}

\vskip 0.3in
]

\printAffiliationsAndNotice{}

\begin{abstract}
Reinforcement learning (RL) in recommendation systems offers the potential to optimize recommendations for long-term user engagement. However, the environment often involves large state and action spaces, which makes it hard to efficiently learn and explore. In this work, we propose a sample-efficient representation learning algorithm, using the standard slate recommendation setup, to treat this as an online RL problem with low-rank Markov decision processes (MDPs). We also construct the recommender simulation environment with the proposed setup and sampling method. 
\end{abstract}

\section{Introduction}
\label{introduction}
Recommender systems aim to find personalized contents based on the learned user preferences, as in collaborative filtering \cite{breese2013empirical,konstan1997grouplens,srebro2004maximum,mnih2007probabilistic} and content-based filtering \cite{van2000using}. A good recommender increases user engagement, and keeps the user interacting within the system. The current popular platforms, like Youtube \cite{covington2016deep}, Spotify \cite{jacobson2016music}, and Netflix \cite{gomez2015netflix}, make extensive use of recommender systems. However, as a user interacts with the system for a longer term, it becomes necessary to keep track of the user dynamics. Traditional methods focus more on myopic predictions, where they estimate the users' immediate responses. Current research has increased to narrate the problem as a Markov decision process \cite{rendle2010factorizing,he2016fusing}, and do long-term planning using reinforcement learning algorithms \cite{shani2005mdp,gauci2018horizon}. However, using RL methods in recommender systems faces an issue regarding the large observation and action space, and doing efficient exploration becomes a harder question. Prior RL methods in recommender systems often overlook exploration, or use $\epsilon$-greedy and Boltzmann exploration \cite{afsar2022reinforcement}. 

In this work, we do representation learning under the low-rank MDP assumption. More specifically, we focus on the case where a user can be learned using a history of observations into a low-dimension representation. And the user dynamics can be modeled as transitions of such representation. Concretely, a low-rank MDP assumes that the MDP transition matrix admits a low-rank factorization, i.e., there exists two unknown mappings $\mu(s')$, $\phi(s, a)$, such that $P(s'|s, a) = \mu(s')^\top\phi(s, a)$ for all $s, a, s'$, where $P(s'|s, a)$ is the probability of transiting to the next state $s'$ under the current state and action $(s, a)$. The representation $\phi(s, a)$ in a low-rank MDP not only linearizes the optimal state-action value function of the MDP \cite{jin2020provably}, but also linearizes the transition operator. Such low-rankness assumption has been shown to be realistic in  movie recommender systems \cite{koren2009matrix}.

Our main contribution is using upper confidence bound (UCB) driven representation learning to efficiently explore in user representation space. It is a practical extension from \textsc{Rep-UCB}, which provides theorical guarantee on efficient exploration under low-rank MDP with sample complexity $O(d^4|\mathcal{A}|^2/(\epsilon^2(1-\gamma)^5))$ \cite{uehara2021representation}. We focus on the case where action is slate recommendation. And under the mild assumption of user choice behavior which we will elaborate in \cref{ass:userchoice}, the combinatorial action space $O(|\mathcal{A}|)$ can be reduced to $O(k|\mathcal{I}|)$ where $k$ is the slate size and $\mathcal{I}$ is the item space.

To evaluate our method systematically, we introduce a recommender simulation environment, \textit{RecSim NG}, that allows the straightforward configuration of an item collection (or vocabulary), a user (latent) state model and a user choice model \cite{mladenov2021recsim}. We describe specific instantiations of this environment suitable for user representation learning, and the construction of our \textsc{Rep-UCB-Rec} learning and optimization methods.

\subsection{Related Works}
\textbf{Recommender Systems}
Recommender systems have relied on collaborative filtering techniques to learn the connection between users and items. Conceptually, they cluster users and items, or embed users and items in a low-dimensional representation \cite{krestel2009latent,moshfeghi2011handling} for further predictions.

For the sake of capturing more nuanced user behaviors, deep neural networks (DNNs) are used in real world applications \cite{van2013deep,covington2016deep}. It naturally follows to be studied as a RL problem, as the user dynamics can be modeled as MDPs. Embeddings are commonly used to learn latent representations from immediate user interactions \cite{liu2020end}. And predictive models are commonly used to improve sample efficiency \cite{chen2021user} and do self-supervised RL \cite{xin2020self,zhou2020s3}.

\textbf{Low-rank MDPs}
Oracle-efficient algorithms for low-rank MDPs \cite{agarwal2020flambe,uehara2021representation} provide sample complexity bounds easing the difficulty of exploration in large state space. Under more restricted setting, such as block MDPs \cite{misra2020kinematic} and $m$-step decodable MDPs \cite{efroni2022provable}, methods are also studied to deal with the curse of dimensionality.

\textbf{Slate-based recommendation and choice models}
Slate recommondation is common in recommender systems \cite{deshpande2004item,viappiani2010optimal,ie2019reinforcement}. Within the context, the complexity within a constructed plate is studied using methods like off-policy evaluation and learning inverse propensity scores \cite{swaminathan2017off}. Hierarchical models are also used for studying user behavior interacting with slates \cite{mehrotra2019jointly}.

The user choice model is linked with the slate recommendation. A common choice model is multinomial logit model \cite{louviere2000stated}. As a good representation of user choice boosts the probability capturing the real-world user behaviors, many areas, such as econometrics, psychology, and operations research \cite{luce2012individual}, have studied it using their own scientific methods. Within the ML community, another popular choice is cascade model \cite{joachims2002optimizing}, as it also captures the fading attention introduced by browsing behavior.

\section{Preliminaries}
\label{preliminaries}
We consider an episodic MDP $\mathcal{M}=\langle \mathcal{S},\mathcal{A},P,r,\gamma,d_0 \rangle$ for slate-based recommendations. In which, a recommender presents a slate to a user, and the user selects zero or one item to consume. Then, the user respond to this consumed item with an engagement measure. The above setup is commonly used and easily extensible to real world applications \cite{ie2019reinforcement}.

Under our setup, the states $\mathcal{S}$ can reflect the ground-truth user states $\mathcal{U}$, which includes static user features (such as demographics), as well as dynamic user features (such as moods). Particularly, the user history interacting with past recommendations plays a key role. This history summarization is usually domain specific, and can capture the user latent state in a partially observable MDP. The state should be predictive of immediate user response (e.g., immediate engagement, hence reward) and self-predictive (i.e., summarizes user history in a way that renders the implied dynamics Markovian).

\begin{figure}[ht]
\begin{center}
\centerline{\includegraphics[width=\columnwidth]{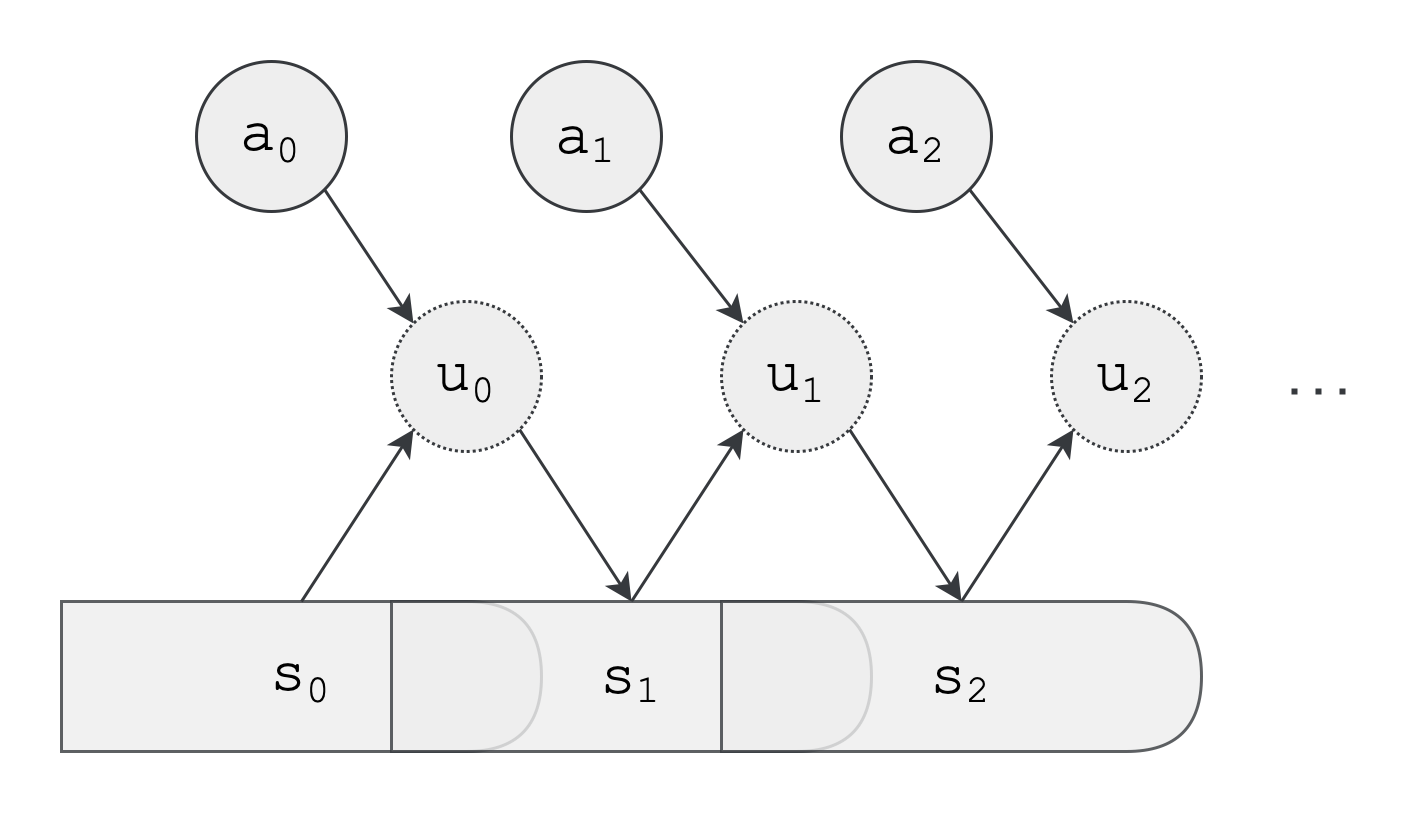}}
\caption{A latent state model captured by low-rank MDP. The states represent the logged user history, including the responses to the recommendations. The $\phi^\star(s,a)$ is a distribution over the latent user representation space $\mathcal{U}$. Note that this is still Markovian model since there is no direct transition between latent states.}
\label{lowrankimg}
\end{center}
\end{figure}

The action space $\mathcal{A}$ is the set of all possible recommendation slates. We assume a fixed set of items 
$\mathcal{I}$ to recommend. Then, an action $a\in\mathcal{A}$ is a subsets $a\subseteq\mathcal{I}$, and $|\mathcal{A}| = {|\mathcal{I}| \choose k}$, where $k$ is the slate size. We assume no constraints, so that each item $i \in \mathcal{I}$ and each slate $a$ can be recommended at each state $s$. Note that we do not account for positional bias within a slate in this work. However, we do note that the effects of ordering within one slate can be learned using offline methods \cite{schnabel2016recommendations}. Because a user may select no item from a slate, we assume that every slate includes a $(k + 1)$-th null item. This is standard in most choice modeling work for specifying the necessary user behaviors induced by a choice from the slate.

The transition $P(s'|s,a)$ represents the probability of user transitioning to $s'$ from $s$ when action $a$ is taken by the recommender. The uncertainty mainly reflects two aspects of a recommender system MDP. First, it indicates how a user will consume a particular recommended item $i \in a$ from the slate, marking the critical role that choice models play in evaluating the quality of a slate. Second, the user state would dynamically transit based on the consumed item. Since the ground truth $P^\star$ is unknown, we need to learn it by interacting with environments in an online manner or utilizing offline data at hand.

The reward $r(s, a)$ usually measures user engagement under state $s$ when recommended with slate $a$. Note that, expectation is more often used to account for the uncertainty introduced by user choice. Without loss of generality, we assume trajectory reward is normalized, i.e., for any trajectory $\{s_h, a_h\}_{h=0}^\infty$, we have $\sum_{h=0}^\infty \gamma^h r(s_h, a_h) \in [0, 1]$. We assume that $r(s,a)$ is known. This assumption largely relies on the success of existing myopic, item-level recommender \cite{covington2016deep}.

The discounted factor $\gamma\in [0,1)$ and initial distribution $d_0\in \Delta(\mathcal{S})$ are also known.

Our goal is to learn a policy $\pi:\mathcal{S}\to\Delta(\mathcal{A})$ which maps from state to distribution over actions (i.e., recommended slates). We use the following notations. Under some probability transition $P$, the value function $V^\pi_P(s)=\mathbb{E}[\sum_{h=0}^\infty \gamma^h r(s_h,a_h)|s_0=s,P,\pi]$ to represent the expected total discounted reward of $\pi$ under $P$ starting at $s$. Similarly, we define the state-action $Q$ function $Q^\pi_P(s, a) = r(s, a) + \gamma \mathbb{E}_{s'\sim P(\cdot|s,a)}V_P^\pi(s')$. Then, the expected total discounted reward of a policy $\pi$ under transition $P$ and reward $r$ is denoted by $V_{P,r}^\pi = \mathbb{E}_{s_0\sim d_0}V_P^\pi(s_0)$. We define the state-action discounted occupancy distribution $d_P^\pi(s,a)=(1-\gamma)\sum_{t=0}^\infty \gamma^t d_{P,t}^\pi(s,a)$, where $d_{P,t}^\pi(s,a)$ is the probability of visiting $(s,a)$ at time step $t$ under $P$ and $\pi$. The state visitation as $d_P^\pi(s)=\sum_{a\in\mathcal{A}}d_P^\pi(s,a)$ Finally, given a vector $a$, $\|a\|_2=\sqrt{a^\top a}$, $\|a\|_B=\sqrt{a^\top B a}$, and $\{c_i\}$ where $i\in\mathbb{N}$ are constants.

We focus on low-rank MDP defined as follows, with normalized function classes.

\begin{definition}(Low-rank MDP)
\label{def:lowrank}
A transition model $P=\mathcal{S}\times\mathcal{A}\to\Delta(\mathcal{S})$ admits a low-rank decomposition with rank $d\in\mathbb{N}$ if there exist two embedding functions $\phi$ and $\mu$ such that \[\forall s,s'\in \mathcal{S}, a\in\mathcal{A}: P(s'|s,a)=\mu(s')^\top\phi(s,a)\] where $\|\phi(s,a)\|_2\leq 1$ for all $(s,a)$ and for any function $f:\mathcal{S}\to[0,1]$, $\|\int\mu(s)f(s)d(s)\|_2\leq \sqrt{d}$. An MDP is low-rank if $P$ admits low-rank decomposition.
\end{definition}

Within the context of recommender systems, low-rank MDPs capture the latent user representation dynamics as shown in  \cref{lowrankimg}. As states are observed and follow Markovian transition, they mainly contain information of the history of user interactions with the recommender. Note that the states are likely to have overlapping information under this definition, and this simplifies the model class of $\mu$ as we will later discuss in \cref{funcapprox}. The actions are the slates presented to the user at each time step. Thus, from a logged history of user and a current slate recommendation, $\phi^\star$ maps $(s,a)$ to the latent representation of user. In reality, the latent space $\mathcal{U}$ would be a compact representation space that contains the most important information that impacts user decisions. Learning such representation and interpreting would be meaningful in real life applications.

In episodic online learning, the goal is to learn a stationary policy $\hat{\pi}$ that maximize $V^{\hat{\pi}}_{P^\star,r}$, where $P^\star$ is the ground truth transition. We can only reset at initial distribution $d_0$, which emphasize the restricted nature of recommender systems and the challenge for exploration. The sampling of state $s$ from visitation $d_P^\pi$ is done by following the \textit{roll-in} procedure: from starting state $s_0\sim d_0$, at every time step $t$, with probability $1-\gamma$ we terminate, otherwise we execute $a_t\sim\pi(s_t)$ and transit to $s_{t+1}\sim P(\cdot|s_t,a_t)$.

\textbf{Function approximation for representation learning.}
\label{funcapprox}
Since $\phi^\star$ and $\mu^\star$ are unknown, we use function classes to capture them. These model classes can be learned using the existing myopic methods for user behavior prediction. 

\begin{assumption} (Realizability)
\label{ass:realize}
We assume to have access to model classes $\phi^\star \in \Phi$ and $\mu^\star \in \Psi$.
\end{assumption}

We assume the function approximators also follow the normalization as $\phi^\star$ and $\mu^\star$, i.e., for any $\phi \in \Phi$ and $\mu \in \Psi$, $\|\phi(s,a)\|_2\leq 1$ for all $(s,a)$, $\|\int\mu(s)f(s)d(s)\|_2\leq \sqrt{d}$ for all $f:\mathcal{S}\to[0,1]$, and  $\int\mu^\top(s') \phi(s,a)d(s')=1$ for all $(s,a)$.

We learn the functions via supervised learning oracle, which should be computationally efficient given the task.

\begin{definition} (Maximum Likelihood Estimator)
The MLE oracle takes a dataset of $(s,a,s')$ tuples, and returns
\[\hat{P} :=(\hat{\mu},\hat{\phi})=\arg\max_{(\mu,\phi)\in \mathcal{M}}\mathbb{E}_{\mathcal{D}_n+\mathcal{D}'_n}[\ln \mu^\top(s')\phi(s,a)].\]
\end{definition}

The assumption above is achievable by the myopic estimators of user behavior models within recommendation context.

\textbf{User choice model.}
\label{ass:userchoice}

We introduce an assumption that is developed to reasonably structuralize the user behavior, and lead to effective reduction on the action space \cite{ie2019reinforcement}.

\begin{assumption} (Reward / transition dependence on selection)
    We assume that $r(s,a)$ and $P(s'|s,a)$ depend only on the item $i\in a$ on slate that is consumed by the user (including the null item). 
    \label{ass:rtds}
\end{assumption}

The original action space under our MDP is ${|\mathcal{I}| \choose k}$ with unordered $k$-sets over $\mathcal{I}$. With large item space, effective exploration is impossible. Luckily, the nature of human preference and structure of recommender systems poses the possibility for reduction. With the prior assumption on user choice setup, where user select zero (null) or one item from slate, we formalize the properties as follows:
    \begin{gather*}
        r(s,a)=\sum_{i\in a}P(i|s,a)r(s,a,i),
        \\P(s'|s,a)=\sum_{i\in a}P(i|s,a)P(s'|s,a,i).
        \\\quad \forall a,a' \text{ containing }i,\quad r(s,a,i)=r(s,a',i)=r(s,i),
        \\P(s'|s,a,i)=P(s'|s,a',i)=P(s'|s,i), \forall a,a'.
    \end{gather*}

Now, notice that the transition can be described as $P(s'|s,a)=\sum_{i\in a}P(i|s,a)P(s'|s,i)$, where $P(i|s,a)$ represents the user's choice on item $i$ given a slate, and $P(s'|s,i)$ is the user state transition when $i$ is consumed. This helps us to redefine uniform action in \cref{main} and obtain complexity bound independent of the combinatorial action space.

Similar to the reward $r(s,a)$, we assume the user choice given a $(s,a)$-pair is known, relying on the success of existing myopic, item-level recommender \cite{covington2016deep}. Thus, the low-rank MDP realizability assumption here is used to describe $P(s'|s,i)$. 

\section{Main results}
\label{main}

We now define uniform action $U(\mathcal{A})$ within the above setup. Remind that the uniform action is used to encourage user transition to some novel states. As user transition is only dependent on consumed item $i$, we use the following definition for sufficient exploration.

\begin{definition}
    The uniform action for a slate recommendation $U(\mathcal{A})$ is defined as the following: 
    \begin{enumerate}
        \item randomly pick an item $i$ from $\mathcal{I}$.
        \item fill the remainder of the slates using the items least-likely be selected by the user.
    \end{enumerate} 
    \label{def:uniaction}
\end{definition}

We further show that under this definition, the effective uniform action space is $O(k|\mathcal{I}|)$.

\begin{lemma}
    The user choose the selected item $i$ with probability at least $1/k$.
    \label{lem:uniaction}
\end{lemma}

We use the basic properties of probability and pigeonhole principle to further arrive at the proposition.

\begin{proposition}
    Efficient exploration action space is achieved in $O(k|\mathcal{I}|)$.
\end{proposition}
\begin{proof} 
    By product rule of probability, the user select an item $i$ with probability at least $1/(k|\mathcal{I}|)$ under the uniform action $U(\mathcal{A})$ as defined above.

    By pigeonhole principle, every $k|\mathcal{I}|+1$ uniform actions lead to at least one duplicate action in this space.
\end{proof}

Note that the above change in distribution of action space definition is introduced by asymmetry of uniform distributions. More specifically, by the principle of indifference, we choose our objective on user state transition instead of naive action space \cite{white2010evidential}. 

\subsection{Algorithm}

\begin{algorithm}[tb]
   \caption{UCB-driven representation learning for recommendation \textsc{(Rep-UCB-Rec)}}
   \label{alg:1}
\begin{algorithmic}
   \STATE {\bfseries Input:} Regularizer $\lambda_n$, parameter $\alpha_n$, Models $\mathcal{M}=\{(\mu,\phi):\mu\in\Psi, \phi\in\Phi\}$
   \STATE Initialize $\pi_0(\cdot|s)$ to be uniform; set $\mathcal{D}_0= \emptyset$, $\mathcal{D}'_0= \emptyset$.
   \FOR{episode $n=1,...,N$}
   \STATE Collect a tuple $(s,a,s',a',\tilde{s})$ with
   \begin{gather*}
       s\sim d^{\pi_{n-1}}_{P^\star},a\sim U(\mathcal{A}),
       \\s'\sim P^\star(\cdot|s,a),a'\sim U(\mathcal{A}),\tilde{s}\sim P^\star(\cdot|s,a)
   \end{gather*}
   \STATE Update datasets \begin{gather*}
   \mathcal{D}_n=\mathcal{D}_{n-1}+{(s,a,s')}, \mathcal{D}'_n=\mathcal{D}'_{n-1}+{(s',a',\tilde{s})}
   \end{gather*}
   \STATE Learn representation via ERM (i.e., MLE) 
   \begin{gather*}
       \hat{P}_n:=(\hat{\mu}_n,\hat{\phi}_n)\\=\arg\max_{(\mu,\phi)\in \mathcal{M}}\mathbb{E}_{\mathcal{D}_n+\mathcal{D}'_n}\left[\ln \sum_{i\in a}\mu^\top(s')P(i|s,a)\phi(s,i)\right]
   \end{gather*}
   \STATE Update empirical covariance matrix 
   \begin{gather*}
       \hat{\Sigma}_n=\sum_{{s,a}\in\mathcal{D}}\hat{\varphi}_n(s,a)\hat{\varphi}_n(s,a)^\top + \lambda_n I\\\text{where } \hat{\varphi}_n(s,a):=\sum_{i\in a}P(i|s,a)\hat{\phi}_n(s,i)
   \end{gather*}
   \STATE Set the exploration bonus
   \begin{gather*}
       \hat{b}_n(s,a):=\min\left(\alpha_n\sqrt{\hat{\varphi}_n(s,a)^\top\hat{\Sigma}_n^{-1}\hat{\varphi}_n(s,a)},2\right)
   \end{gather*}
   Update policy \begin{gather*}
       \pi_n = \arg\max_\pi V^\pi_{\hat{P}_n,r+\hat{b}_n}
   \end{gather*}
   \ENDFOR
\end{algorithmic}
\end{algorithm}

The algorithm is based on $\textsc{Rep-UCB}$ \cite{uehara2021representation}, with specified definition on state space $\mathcal{S}$, action space $\mathcal{A}$, the data collection process, and the uniform action $U(\mathcal{A})$ under the context of recommender system.

A state $s$ consist of static user features and a history of past recommendations and user responses. An action $a$ provides a slate recommendation. During the data collection process, for every iteration, we do one rollout under current policy $\pi$. We assume the states sampled from the initial distribution $d_0$ follows the prior \cref{ass:realize} on model classes. The sampling procedure for $s\sim d_{P^\star}^{\pi_{n-1}}$ begins with $s_0\sim d_0$. At every time step $t$, we terminate with probability $1-\gamma$, or execute $a_t\sim \pi(s_t)$ and observe $s_{t+1}\sim P^\star(\cdot|s_t,a_t)$. We sample uniform action $U(\mathcal{A})$ defined by \cref{def:uniaction}. It means the recommender would do recommendations based on planning until termination, then do two uniform recommendations by the end of data collection process. We collect the tuple $(s,a,s',a',\tilde{s})$ and update datasets.

Representation learning, building the empirical covariance matrix, and calculating the bonus are done sequentially after datasets updates. The final step within the episode is planning using the learned estimation of transition with added exploration bonus.

\subsection{Analysis}
The PAC bound for $\textsc{Rep-UCB}$ \cite{uehara2021representation} provides us a good starting point. Our anlaysis focuses on reducing the naive combinatorial action space $|\mathcal{A}|$ to be polynomial in slate size $k$ and item space size $|\mathcal{I}|$.

\begin{theorem} (PAC Bound for \textsc{Rep-UCB-Rec})
Fix $\delta \in (0,1)$, $\epsilon \in (0,1).$ Let $\hat{\pi}$ be a uniform mixture of $\pi_1,...,\pi_N$ and $\pi^\star:=\arg\max_\pi V^\pi_{P^\star,r}$ be optimal policy. Set the parameters as follows:
\[\alpha_n = O\left(\sqrt{(k|\mathcal{I}|+d^2)\gamma\ln(|\mathcal{M}|n/\delta)}\right),\]\[ \lambda_n=O(d\ln(|\mathcal{M}|n/\delta)),\]with probability at least $1-\delta$, we have \[V^{\pi^\star}_{P^\star,r}-V^{\hat{\pi}}_{P^\star,r}\leq\epsilon.\]
The number of collected samples is at most \[O\left(\frac{d^4k^2|\mathcal{I}|^2\ln(|\mathcal{M}|/\delta)^2}{(1-\gamma)^5\epsilon^2}\cdot \nu\right)\] where \begin{gather*}
    \nu:= O\left(\ln\left(\frac{\iota}{\delta}\ln^2\left(1+\iota\right)\right)
    \cdot \ln^2\left(1+\iota\right)\right)
\end{gather*} and \[\iota:=\frac{d^4k^2|\mathcal{I}|^2\ln(|\mathcal{M}|/\delta)^2}{(1-\gamma)^5\epsilon^2}.\]
\end{theorem}

\begin{proof}
    The upper bound dependency on $|\mathcal{A}|$ is introduced by the importance weighing of the policy $\pi$ to uniform action, where $\max_{(s,a)}\frac{\pi(a|s)}{u(a)}\leq |\mathcal{A}|$. Under our definition of $U(\mathcal{A})$ under slate recommendation context, we naturally change the upper bound to $k|\mathcal{I}|$. Note that this dependency is then interchangeable throughout the proof.
\end{proof}

\section{Simulations}

We discuss the simulation environment setup and the algorithm construction in this section. The simulation uses \textit{Recsim NG} \cite{mladenov2021recsim}. A graph illustration is shown in \cref{flow}.

\subsection{Simulation environment}

\textbf{Item class.} A static set of items are sampled at the beginning of the simulation. Each item is represented by a $T$-dimensional vector $\mathbf{i}\in [-1,1]^T$, where each dimension represents a topic. Each item has a length $l(\mathbf{i})\in [0,1]$ (e.g. length of a video, music or an article), and a quality $q(\mathbf{i})\in [-1,1]$ that is unobserved from the user. 

\textbf{User interest.} Users $\mathbf{u}\in U$ have various degrees of interests in each topics. Each user $\mathbf{u}$ is represented by an interest vector $\mathbf{u}\in [-1,1]^T$. The user interest towards a certain item is calculated by inner product $\mathbf{i}^\top \mathbf{u}$. The user interest vector $\mathbf{u}$ and the mechanism of user interest towards items are unobserved to recommenders.

\textbf{User choice.} Given a slate of $k$ items, a user choose to consume one item from the slate with \[P(\mathbf{i}|a,\mathbf{u} )=\frac{e^{\mathbf{i}^\top \mathbf{u}}}{\sum_{\mathbf{j}\in a} e^{\mathbf{j}^\top \mathbf{u}}}.\]
This choice model is called multinomial logit model \cite{louviere2000stated}. For the null item (no choice), the item is simply represented by a $T$-dimensional zeros vector with length and quality zeros.

\textbf{User dynamics.} The internal transition of user interest vector allows the environment to capture Markovian transition and allow RL methods to do meaningful planning. After the consumption on item $\mathbf{i}$ at time step $t$, a user $\mathbf{u}$ follows
\[\mathbf{u}_t = c_0 \mathbf{u}_{t-1} + c_1 q(\mathbf{i})(\mathbf{i}-\mathbf{u})+ \epsilon\] where $\epsilon\sim\mathcal{N}(0,c_3).$ The constants are for normalization. Under this update, the user transits to favor the topics in $\mathbf{i}$ more if the quality of $\mathbf{i}$ is positive.

\textbf{Reward.} The reward is reflected by the user consumption time of a chosen item $\mathbf{i}$. It is linear with respect to the length and the user interest of item $\mathbf{i}$.

\subsection{Algorithm construction}

 The state is a history of length $h$ that the user interacts with the recommender. The history contains the slate recommendations and user responses at each time step $t-h,...,t-1$. We construct the sampling procedure introduced by $\textsc{Rep-UCB-Rec}.$ At each episode, the recommender follows the current learned policy and observe user responses. Until the rollin procedure terminates, the recommender further does two step uniform recommendation. The supervised learning for representation is done in an offline manner, where we combine two model classes and train them together to make predictions from $(s,a)$ to $s'$. Note that by definiton of states, we only predict the user next response towards current action, and slide up one time step the time window of history. It eases the complexity of large state space to $O(k)$ and reflects on the special structure behind recommender systems. After calculating the emperical covariance matrix and exploration bonus, we utilize a costomized simulator where transition and reward are estimated. Standard policy gradient methods are used to calculate the updated policy within the this environment under $\hat{P}$ and $(r+\hat{b})$.

 \section{Conclusion}
 In this work, we propose a sample-efficient representation learning algorithm, using the standard slate recommendation setup, to treat this as an online RL problem with low-rank Markov decision processes (MDPs). We show that the sample complexity for learning near-optimal policy is $O(d^4k^2|\mathcal{I}|^2/\epsilon^2(1-\gamma)^5)$ where $k$ is the slate size and $\mathcal{I}$ is the item space. We further show the detailed construction of a recommender simulation environment with the proposed setup and sampling method.

\bibliography{main}
\bibliographystyle{icml2023}

\newpage
\appendix
\onecolumn
\section{Graph illustration of the simulation environment.}
\label{flow}
\begin{figure}[ht]
\begin{center}
\centerline{\includegraphics[width=\columnwidth]{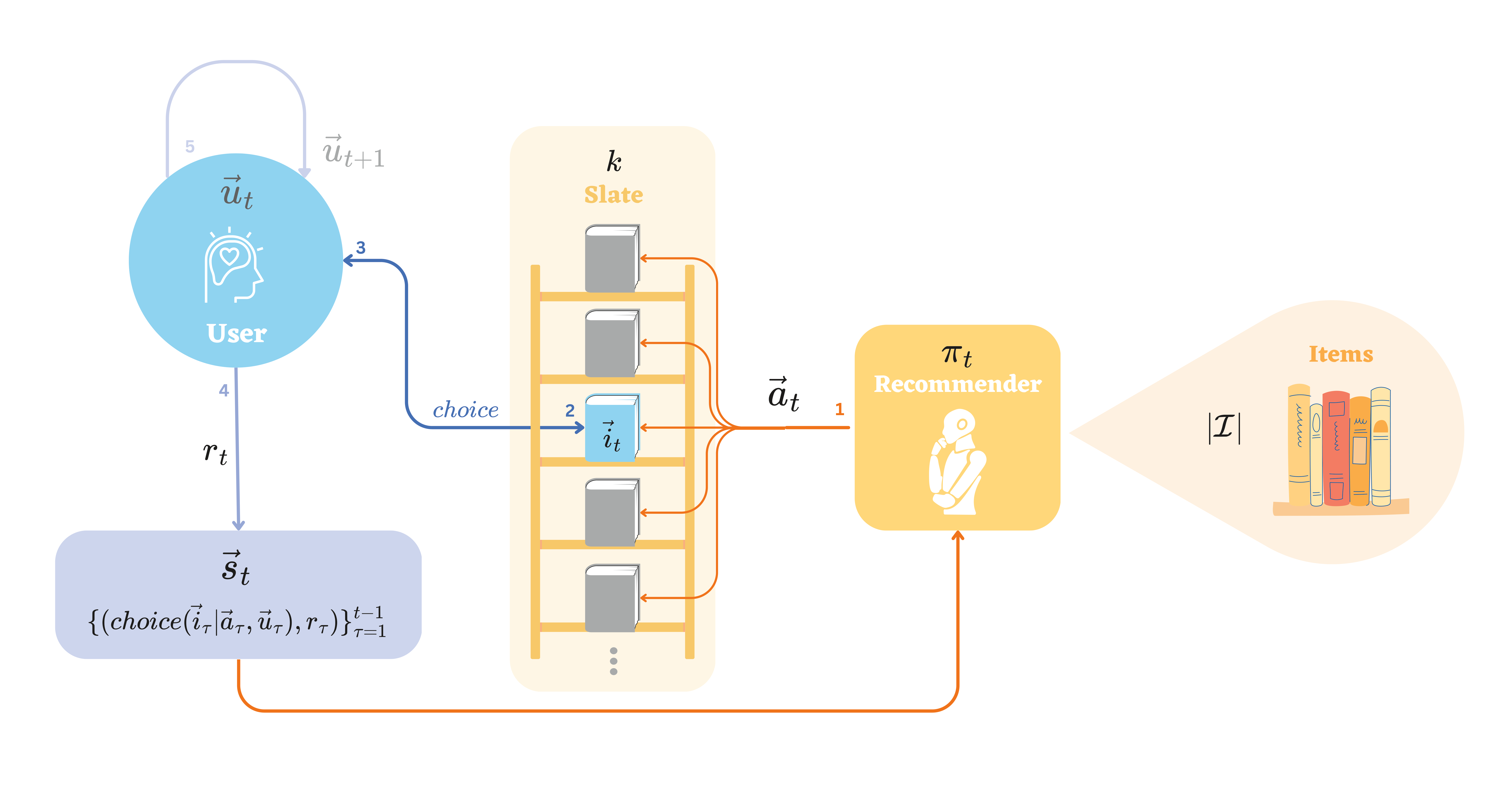}}
\caption{This flowchart describes the sequential decision making process of the recommender system setup we focus on. More specifically, the user choose an item $\vec i_t$ from the slate-based recommendation, transit to a new user internal state $\vec u_{t+1}$, and emit a reward $r_t$ and new observable state $\vec s_{t+1}$. The recommender learns to use the observed state to predict user internal state and plan with exploration and exploitation.}
\label{lowrankimg}
\end{center}
\end{figure}

\end{document}